\documentstyle[eqsecnum,epsf,prd,aps]{revtex}
\begin{document}
\draft
\twocolumn[
\title{Non-linear instability of Kerr-type Cauchy horizons}
\author{Patrick R. Brady and Chris M. Chambers}
\address{Department of Physics, The University, Newcastle Upon Tyne NE1 7RU}
\date{\today}
\preprint{NCL94-TP13}

\maketitle

\typeout{ABSTRACT}

\begin{abstract}
\widetext
Using the general solution to the Einstein equations on intersecting null
surfaces developed by Hayward, we investigate the
non-linear instability of the Cauchy horizon inside a realistic black hole.
Making  a minimal assumption about the free gravitational data allows us to
solve the field equations along a null surface crossing the Cauchy Horizon. As
in the spherical case, the
results indicate that a diverging influx of gravitational energy, in concert
with an outflux across the CH, is responsible for the singularity.
The spacetime is asymptotically Petrov type N,  the same algebraic type as
a gravitational
shock wave.  Implications for the continuation of spacetime through the
singularity are briefly discussed.
\end{abstract}
\pacs{PACS:  97.60Lf, 04.70-s, 04.20Dw}
]
\narrowtext

\typeout{INTRODUCTION}

\section{Introduction}

What are the generic features of gravitational collapse?  We do not, as yet,
have a complete answer to this question,  however most relativists are
confident
 that gravitational collapse leads to the formation of black holes (at least
within the astrophysical context).  Furthermore it is plausible that the
external field of such a black hole settles down to a member of the
Kerr-Newman family, since these are the unique stationary solutions of the
electrovac Einstein equations (see for example~\cite{Wald}).  Externally,
deviations from these solutions are expected to die away with an inverse
power-law in advanced time leaving only the mass,  charge and angular momentum
observable.  A question which one might hope to answer is whether such a
property continues to hold inside the black hole -- that is, does the internal
geometry also approach Kerr-Newman form?  It seems
not~\cite{PI:90}-\cite{Ori:92}.

The known exact black-hole solutions possess Cauchy horizons -- null
hypersurfaces which are the boundary of the future domain of dependence  for
Cauchy data of the collapse problem.  These horizons exhibit highly
pathological behaviour; small time-dependent perturbations originating outside
the black hole
undergo an infinite gravitational blueshift as they evolve towards the Cauchy
horizon~(CH).  This blueshift of infalling radiation gave the first indications
that
these solutions, which so well describe the exterior geometry at late times,
may not describe the generic internal structure.  Penrose~\cite{Penrose1}
pointed
this out more than twenty five years ago, and since then linear perturbations
have
been analysed in detail~\cite{Penrose2}.  These observations led
to the conjecture that a scalar
curvature singularity would form either {\em at} or {\em before} the
CH once back-reaction was accounted for.

Poisson and Israel~\cite{PI:90}, generalizing work of Hiscock~\cite{Hiscock},
have shown that a scalar curvature singularity forms along the
CH of a charged, spherical black hole in a simplified model. This singularity
is
characterised by the exponential divergence of the mass function with advanced
time.  The key ingredient producing this tremendous growth of curvature is the
blueshifted radiation flux along the CH,  although it is also necessary that
some transverse energy flux be present.  In~\cite{PI:90} it was
argued that the
physics underlying the analysis was sufficiently general that similar
results should hold for generic collapse (i.e.  upon relaxation of the
assumption of spherical symmetry).  Further calculations support the
conjecture that the singularity  inside a generic black hole
is null~\cite{Ori:92,Bradyetal}.

The aim of this paper is to present a detailed analysis of the CH inside a
non-spherical black hole by taking advantage of the recent result due to
Hayward~\cite{Hayward}. He showed how to obtain the general solution of the
Einstein equations on a pair of intersecting null surfaces.  We begin by
recasting the results of Poisson and Israel~\cite{PI:90} in a suggestive form
which more closely resembles the approach taken to the general problem.
Our purpose in section~\ref{section1} is to stress the main points which must
be
taken as
assumptions in the later calculations.  We also feel that the analysis
highlights the non-linear nature of the effect and the merely precipitous role
played by the outflux in this model~\cite{PI:90}.
With this preliminary review of the
mass-inflation phenomenon out of the way, section~\ref{section2} begins the
analysis of the CH singularity inside a realistic black hole.   The assumptions
which we make about the nature of the CH are discussed,  and following
Hayward~\cite{Hayward} we formally integrate the first order Einstein equations
of appendix~\ref{appendixa}.  Using these results we can show that the
divergences which arise are integrable,  and obtain the leading behaviour of
all quantities of interest near to the CH. The asymptotic expressions for
the Weyl scalars show that
the spacetime is Petrov-type N near the CH, the same algebraic type as a
gravitational shock wave.  These results (which
are also suggested by the previous work~\cite{Bradyetal}) once again raise
questions about the classical continuation of spacetime through the
singularity.

The equations and curvatures are relegated to the appendices in an effort to
maintain the clarity of the presentation.  Appendix \ref{appendixa} gives a
summary of the notation and lists the dual null Einstein
field equations in their first order form, as derived by
Hayward~\cite{Hayward}.
Appendix~\ref{appendixc} lists the components of the
Riemann tensor necessary to analyse the algebraic type of the spacetime.


\section{Spherical black hole interiors}\label{section1}


In this section we review the mass-inflation phenomenon in the context of
charged, spherical black holes as studied by Poisson and Israel~\cite{PI:90}.
The presentation differs slightly from that in the
literature~\cite{PI:90,Ori:91} and closely parallels the method used later to
discuss the non-spherical problem.  In this way the limitations of the
later analysis,  and of the approximations used should be more apparent.

The physics behind mass-inflation is relatively simple.  The CH inside a
Reissner-Nordstr\"om black hole is a null hypersurface corresponding to
infinite
external advanced time.  Time-dependent perturbations which originate in the
external universe are gravitationally blueshifted as they propagate inwards
near to the CH.  Thus a charged black hole which deviates only slightly from
Reissner-Nordstr\"om in the exterior is expected to have a barrier of
radiation,
with an exponentially diverging energy  density,
streaming along parallel to the CH.  Generically some of this ingoing radiation
scatters off the
gravitational potential inside the black hole, leading to a flux of energy
crossing the CH~\cite{Drozetal}.  It is the non-linear gravitational
interaction of
these two fluxes of radiation which generates a divergence of the local mass
function.
 It is important to realise that the gravitational blueshift of time-dependent
perturbations is the key ingredient producing this {\em mass-inflation}.


\subsubsection*{The Poisson-Israel model}\label{section1.2}


It is convenient to use null coordinates on the ``radial'' two
spaces so that the spherical line element is
   \begin{equation}
   ds^2 = -2e^{-\lambda}dudv + r^2(dx^2 + \sin^2xdy^2)\: ,\label{2.1}
   \end{equation}
where $\lambda=\lambda(u,v)$, $r=r(u,v)$ and the coordinates  are such that
$u$ is a
retarded time and $v$ an advanced time.  The
stress-energy tensor for a radial electromagnetic field is
   \begin{equation}
   E{_\mu}^{\nu} = q^2/8\pi r^4  \:\mbox{\rm diag}(-1,-1,1,1)\; , \label{2.2}
   \end{equation}
where $q$ is the electric charge on the black hole.
Poisson and Israel used crossflowing
null dust to model the perturbations of the geometry,  arguing that the large
blueshift near to the Cauchy horizon should make the Isaacson~\cite{Isaacson}
effective stress energy desciption valid for the ingoing radiation.  They also
pointed out that the nature of the outflux is not important,  its only purpose
is to initiate the contraction of the Cauchy horizon.  The stress-energy tensor
is therefore
	\begin{equation}
	T_{\mu\nu} = {\rho}_{\rm in} l_{\mu} l_{\nu} + {\rho}_{\rm out}
	n_{\mu} n_{\nu}
	\: , \label{2.3}
	\end{equation}
where $l_{\mu} = -\partial_{\mu} v$ and $ n_{\mu} =  -\partial_{\mu}
u$ are
radial null vectors pointing inwards and outwards respectively, and,
$\rho_{in}$ and $\rho_{out}$ represent the energy
densities of the inward and outward fluxes.  Each term in Eq.~(\ref{2.3}) is
independently conserved so that
	\begin{equation}
	\rho_{\rm in} = \frac{L_{\rm in}(v)}{4\pi r^2} \: ,
	\mbox{\hspace{.25in}}
	\rho_{\rm out} = \frac{L_{\rm out}(u)}{4\pi r^2} \: .
	\label{2.4}
	\end{equation}
The functions $L_{\rm in}(v)$ and $L_{\rm out}(u)$ are determined by the
boundary conditions, however, it is important to note that they have no direct
operational meaning since they depend on the parametrization of the null
coordinates.

\begin{figure}
\leavevmode
\hbox{\epsfxsize=8cm \epsfysize=8cm {\epsffile{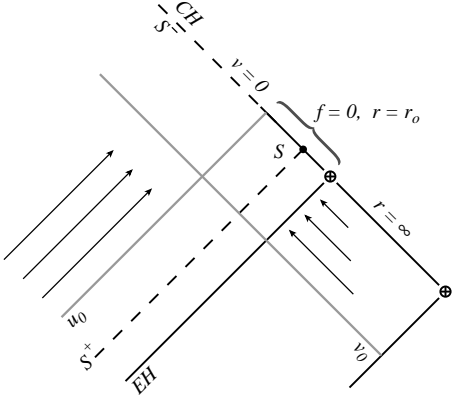}}}
\caption{\label{figure1}A spacetime diagram showing the spherical
Poisson-Israel
model.
The influx (outflux) of lightlike dust is turned on at advanced (retarded)
time $v_0$ ($u_0$).  The surface $S^-$ coincides with the Cauchy horizon, CH.
$S^+$ is parallel to the event horizon, EH, and intersects the Cauchy horizon
at
S, on which $f=0$ and $r = r_o$.}
\end{figure}

The field equations can now be written in a first order form by defining the
extrinsic fields
	\begin{eqnarray}
	&&\theta := r^{-2}\; (r^2)_{,v} \label{101}\; , \\
	&&\widetilde{\theta} := r^{-2}\; (r^2)_{,u} \label{102}\; , \\
	&&\nu := \lambda_{,v} \label{103}\; , \\
	&&\widetilde{\nu} := \lambda_{,u} \label{104}\; ,
	\end{eqnarray}
where a comma denotes partial differentiation.
Along $S^{-}$ there are two evolution equations
	\begin{eqnarray}
	(r^2 \, \theta)_{,u} &=& {e^{-\lambda} \over r^2} ( { q^2  }  -  r^2 )
	\label{105} \\
	(\theta - 2\nu)_{,u} &=& { e^{-\lambda}  \over  { r^4}} ( r^2 - 3 q^2 )
	\label{106}\; ,
	\end{eqnarray}
and a focussing equation which describes the behaviour of the
null cones with vertices at $r=0$:
	\begin{eqnarray}
	(r^2\, \widetilde{\theta})_{,u} + r^2(\widetilde{\nu} -
	\widetilde{\theta}/2)\, \widetilde{\theta} &=&
		-2 L_{\rm out}(u) \label{108} \; .
	\end{eqnarray}
The complete set of Einstein equations,  including those which hold on $S^+$,
may be obtained from the above equations (\ref{105}),
(\ref{106}) and (\ref{108})
via the symmetry operation
	\begin{eqnarray}
	(u;\widetilde{\theta},\widetilde{\nu},\theta,\nu) &\rightarrow
	&
	(v;\theta,\nu,\widetilde{\theta},\widetilde{\nu})\; , \nonumber\\
	\label{104a}\\
	L_{\rm out} &\rightarrow& L_{\rm in}.  \nonumber
	\end{eqnarray}
It is convenient to imagine that the inflow (outflow) is turned on at some
advanced (retarded) time $v=v_0$ ($u=u_0$).  In the pure inflow regime the
spacetime is described by a charged Vaidya solution
    \begin{equation}
    ds^2  =  dw ( 2 dr - f  dw)  +  r^2 ( dx^2  +  \sin^2x  d y^2 ),
    \label{2.13}
    \end{equation}
where $f$ is given by
    \begin{equation}
    f  =   1  -  { 2 m ( w )  \over  r}   +   { q^2   \over   r^2 } \; ,
    \label{2.14}
    \end{equation}
and $w$ is the {\em standard} external advanced time coordinate.  In particular
it is infinite both on future null infinity and the CH.  The only non-zero
component of the stress-energy tensor is
	\begin{equation}
	T_{ww} = \frac{1}{4\pi r^2} \frac{dm}{dw} \; . \label{2.15}
	\end{equation}
In the outflow region the solution is of the same form with advanced replaced
by
retarded time.  The structure of the spacetime is shown in Fig.~\ref{figure1}.

\subsubsection*{The solution}

As mentioned above our treatment differs from that in the
literature~\cite{PI:90,Ori:91}.
We consider the solution to the Einstein equations only on the two
intersecting null surfaces $S^- $ and $S^+ $.  By choosing $S^- $ to coincide
with the CH we can show that a scalar curvature singularity must be present on
the CH in the cross-flow region.

We first integrate the equations on the CH.  This is readily
achieved by making an appropriate choice of the retarded time coordinate $u$,
in
(\ref{2.1}), such that $\widetilde{\nu} = \widetilde{\theta}/2$ along $S^- $.
Then (\ref{104}),
(\ref{108}) and (\ref{102}) can  be integrated, in that order, to give
	\begin{eqnarray}
	e^{\lambda_-} &=& r_- \; ,  \label{2.16} \\
	\widetilde{\theta}^- &=& \left( r_o^2\, \widetilde{\theta}^o -
	2\int^u_0\!\! d\tilde{u} \;
	L_{\rm out}(\tilde{u})
	\right) / r_-^2\; ,  \label{2.17}\\
	r^2_- &=& r^2_o + r_o^2\, \widetilde{\theta}^o \, u -
	2\int^u_0\!\!d\hat{u}\;
	\int^{\hat{u}}_0\!\!d\tilde{u}\;  L_{\rm out}(\tilde{u})  \; .
	\label{2.18}
	\end{eqnarray}
The (sub-) super-script $(-)$ is used to indicate that these
equalities hold only on $S^- $,  while an $(o)$ indicates the value of
the function on the two-sphere $S = S^- \cap S^+$.  Of the remaining two
equations we only need (\ref{105}) which gives
	\begin{equation}
	r_-^2\, \theta^- = r_o^2\, \theta^o + \int_0^u \!\!  d\tilde{u}\;
	 \left( { q^2   \over  \tilde{r}^3 }  -  { 1
	\over
	\tilde{r}}\right)   \; . \label{2.19}
	\end{equation}
 Eq.~(\ref{106}) could be integrated in the same way.

The same construction works on $S^+ $  (we will use this as our
primary tool in the non-spherical case)  however it is more convenient to work
with the exact solution (\ref{2.13}) to determine $\widetilde{\theta}^0$,
$\theta^0$ and
$r_0$  here.

\subsubsection*{Assumptions}

The two essential features of this model will be adopted in the non-spherical
case  with  only slight modification. We wish to emphasise them at this point.

The first assumption is that there should exist a stationary portion of the CH
in the spacetime.  This is achieved by turning on the outflux at some retarded
time ($u_0$) inside the event horizon of the black hole (see Fig.
\ref{figure1}),  thus
set
	\begin{equation}
	L_{\rm out}(u) = \beta H(u-u_0) \; ,\label{2.20}
	\end{equation}
where $\beta$ is a constant,  and $H(u)$ is a Heaviside step function.
This is probably the most contentious issue in the Poisson-Israel~\cite{PI:90}
and subsequent analyses~\cite{Ori:92}.  It presupposes that the CH begins at
finite radius,  and essentially assumes that the singularity at the CH will be
null.  Some arguments have been advanced to suggest that this issue is
important~\cite{Yurt},  however no convincing evidence has emerged to refute
its
correctness.  Indeed preliminary numerical results~\cite{BGS}
support this assumption,  in contrast to the claim in~\cite{Gnedin}.

The second assumption pertains to the influx of radiation; it is taken to
decay with an inverse power law of advanced time.  This was initially based on
an extrapolation of results due to Price~\cite{Price} to the event horizon of
the black hole.  Recently it has been verified numerically~\cite{Priceetal}.
This decay is correctly reproduced by the ansatz
	\begin{equation}
	m(w) = m_o - { \alpha  r_o \over (p-1) }
              ( \kappa_o  w )^{-(p-1)} \; ,  \label{2.21}
	\end{equation}
where $\kappa_o$ is the surface gravity of the inner horizon,  $\alpha$ is a
dimensionless constant which depends on the luminosity of the collapsed star,
and $p \ge 12$  is an integer  (the numerical value derives from Price's
analysis which shows that the radiative tail of the collapse decays as an
inverse power  $p \ge 4l+4$
of advanced time,  where $l$ is the multipole order of the perturbation).  Both
linear~\cite{Penrose2,Chandra} and non-linear~\cite{Ori:92} perturbation
analyses suggest that this inverse power-law decay of perturbations is also
correct
near the CH inside the black hole.

\subsubsection*{Mass inflation}

Along $S^+ $ the radius obeys the first order equation
	\begin{equation}
	\frac{dr}{dw} = \frac{1}{2}\left( 1 - \frac{2m(w)}{r} +
	\frac{q^2}{r^2} \right)  \label{2.22}
	\end{equation}
where $m(w)$ is given by (\ref{2.21}).  We are interested in calculating
	\begin{equation}
	\theta^+ = 2 r^{-1}_+ \frac{dr_+}{dw}\frac{dw}{dv} \label{2.23}
	\end{equation}
where $v$ is the coordinate which appears in (\ref{2.1}) chosen so that
$\lambda
=
\ln r$ along $S^+ $.  It is a straightforward matter to calculate $dw/dv$
provided one notes that Eqs. (\ref{101})-(\ref{108}) are form invariant under
functional rescalings of the coordinates $u$ and $v$.  Therefore  $r$ also
satisfies
	\begin{equation}
	\frac{d^2{}}{dw^2}(r^2) + \frac{d}{dw} \left( \ln \left[\frac{dw}{dv}
	\right]
	\right)  \frac{d}{dw}(r^2) = -2 \frac{dm}{dw} \label{2.24}
	\end{equation}
on $S^+ $ in view of (\ref{104a}) and (\ref{108}).  Solving (\ref{2.22}) for
$r$,  (\ref{2.24})
can be
integrated to get
	\begin{equation}
	v \simeq - e^{-\kappa_o w} ( 1 + \ldots)\label{2.25}
	\end{equation}
as $w \rightarrow \infty$.  Thus
	\begin{equation}
	r_+ \simeq r_o +  \frac{\alpha (-\ln|v|)^{-p+1} }{(p-1) \kappa_o}
	\left\{
	1 + (p-1)(-\ln|v|)^{-1} + \ldots \right\}
	\label{2.26}
	\end{equation}
and
	\begin{equation}
	\theta^+ \simeq \frac{2\alpha (-\ln|v|)^{-p}}{r_o \kappa_o\, v} \left\{
	1 + p(-\ln|v|)^{-1} + \ldots\right\}  \label{2.27}
	\end{equation}
as $v\rightarrow 0$.  This result means that, while the radius of the
two-sphere
$S$ is finite,  $\theta^o$ is actually infinite (in these coordinates).

Using the invariant definition of the mass in a spherical geometry
	\begin{equation}
	1 - \frac{2m(x^{\alpha})}{r} + \frac{q^2}{r^2} :=
	g^{\alpha\beta}r_{,\alpha} r_{,\beta} = -\frac{ r^2 e^{\lambda} \theta
	\widetilde{\theta}}{2}
	\label{2.28}
	\end{equation}
we know that $\theta^+ \widetilde{\theta}^+ \rightarrow 0$ as $v\rightarrow 0$,
which
means
that $\widetilde{\theta}^o = 0$.  Substituting (\ref{2.20}) into (\ref{2.17})
and
using $\widetilde{\theta}^o = 0$
	\begin{equation}
	r^2_-\, \widetilde{\theta}^- = -2 (u-u_0) \beta  H(u-u_0) \; .
\label{2.29}
	\end{equation}
The radius of the CH is then
	\begin{equation}
	r^2_- = r^2_o - (u-u_0)^2 \beta H(u-u_0) \; ,\label{2.30}
	\end{equation}
and finally $\theta^-$ is given by (\ref{2.19}).   The  square of the Weyl
curvature  on
the CH is
	\begin{equation}
	{\cal C}^- = \left.
	C_{\alpha\beta\gamma\delta}C^{\alpha\beta\gamma\delta}\right|_{S^-}
	= 12 \left( \frac{1}{r^2_-} - \frac{q^2}{r^4_-} +
	{r_- \over 2} {\theta^- \widetilde{\theta}^-}
	\right)^2 \; , \label{2.31}
	\end{equation}
from which we can immediately see the reason for the Cauchy horizon
singularity.
The product $\theta^- \widetilde{\theta}^- =0$ at $S$ (see Fig.~\ref{figure1})
and hence
${\cal C}^-$ is finite,
however, if $\beta$ is non-zero ${\cal C}^-$ jumps to infinity at $u=u_0$ when
the outflux is turned on [see Eq. (\ref{105})].  Furthermore Eq. (\ref{2.30})
shows
that the radius remains finite for some amount of retarded time,   so it must
be the mass function which becomes infinite at $u=u_0$.

This analysis emphasises that the nature of the outflux is largely
irrelevant,  it simply serves to initiate the contraction of the CH.  It is
this contraction which is at the root of mass-inflation~\cite{PI:90}.

\section{Non-spherical black holes}\label{section2}

Charged spherical black holes have served as a useful model in which to
investigate the non-linear instability of the CH inside a black hole,  however,
the question needs to be addressed in the more realistic non-spherical context.
 Here one expects that a black hole settles down to a member of the
Kerr-Newman family at late times.  (Our analysis focuses
on such a situation when there is no electromagnetic field.)

The Kerr solution can be written as
	\begin{eqnarray}
	ds^2 \! &=&  \! -\frac{\Delta}{\rho^2}(dw \! - \! a d\phi
	\sin^2\theta)^2
		+ \frac{\sin^2\theta}{\rho^2} ( [r^2 + a^2] d\phi
		\! - \! a dw)^2 \nonumber\\
	&& \ \ + 2 dr (dw \! - \! a d\phi \sin^2\theta) + \rho^2d\theta^2
	\label{3.1}
	\end{eqnarray}
where $\Delta = r^2 - 2mr + a^2$,  $\rho^2 = r^2 + a^2 \cos^2 \theta$ and $w$
is standard advanced time. The CH, located at $r= m - \sqrt{m^2 - a^2}$ and
$w=\infty$, is
a stationary null hypersurface  --  its lightlike generators have zero rate of
expansion.  Moreover,  these generators are shear free since Raychaudhuri's
equation implies that shear produces contraction.  This is not a generic
situation;  a linear perturbation analysis shows that there is usually a flux
of backscattered radiation crossing the CH, along with the blueshifted influx
parallel to it~\cite{Chandra}.  Based on this observation and using the
spherical case as a guide,  we will show that the CH is the locus of a scalar
curvature singularity.  The leading divergence may be associated with
propagating modes of the gravitational field,  manifesting itself in the form
of a diverging Weyl curvature of Petrov type-N.

\subsubsection*{Formalism}

We employ the dual null formalism of Hayward~\cite{Hayward} which is based on a
2+2 null decomposition of spacetime~\cite{dInverno}.  In terms of coordinates
$u$ and $v$ which label the intersecting null hypersurfaces that foliate the
spacetime,  the line element can be written as
	\begin{equation}
	ds^2  =   -   2 e^{-\lambda} du dv  + (s_a s^a) du^2   +   2 s_{a}
	dx^{a} du
	+   h_{ab} dx^{a} dx^{b}                  \label{3.2}
	\end{equation}
where the lower-case latin indices range over $\{ 1,2 \}$. There remains
freedom to rescale $u$ and $v$,  and to transform the coordinates $x^a$ so that
$s^a$ vanishes on a chosen $v = $~constant surface, we take
this to be the CH located at $v=0$  (see Fig.~\ref{figure2}).  $h_{ab}$ is the
{\em 2-metric},  $s^a$ a shift {\em
2-vector}
and $\lambda$ a scaling {\em function} on $S(u,v)$,  the spatial 2-surface
which
is the intersection of the null surfaces labelled by $u$ and $v$.  The 2-metric
can be
decomposed into a conformal factor $\Omega$ and a conformal {2-metric} $k_{ab}$
so that
	\begin{equation}
	\Omega = \sqrt{h}\; , \ \ \ \  k_{ab} = \Omega^{-1} h_{ab} \; .
	\label{3.3}
	\end{equation}
Thus $k_{ab}$ has unit determinant.  Each of the quantities defined here depend
on all four variables ${u,v, x^a}$.

The vacuum Einstein equations (\ref{E1})-(\ref{E19}) are written in a first
order form
in
terms of the fields
	\begin{eqnarray}
	\hbox to 0.19in{$\displaystyle \theta$\hfill}
	&=& \hbox to .75in{$\displaystyle \Omega^{-1} {\cal L}_{l}
	 \Omega$, \hfill}
	 \mbox{\hspace{.5in}}
	\hbox to .19in{$\displaystyle {\widetilde{\theta}}$\hfill}
	= \Omega^{-1} {\cal L}_{n} \Omega \label{3.4}\\
	\hbox to 0.19in{$\displaystyle \sigma_{ab}$\hfill}
 	&=& \hbox to .75in{$\displaystyle \Omega \,{\cal L}_{l} k_{ab}$,\hfill}
 	\mbox{\hspace{.5in}}
	\hbox to .19in{$\displaystyle \widetilde{\sigma}_{ab}$\hfill}
	= \Omega \, {\cal L}_{n} k_{ab} \label{3.5}\\
	\hbox to 0.19in{$\displaystyle \nu$\hfill}
 	&=& \hbox to 0.75in{$\displaystyle {\cal L}_{l} \lambda$,\hfill}
  	\mbox{\hspace{.5in}}  \hbox to 0.19in{$\displaystyle
	\widetilde{\nu}$\hfill} =
	{\cal L}_{n} \lambda
	\label{3.6}\\
	&&\mbox{\ \ \ }\omega_a = \frac{1}{2} e^{\lambda}\, \Omega k_{ab} \,
	{\cal L}_{l}
	s^b \; ,
	\label{3.7}
	\end{eqnarray}
where $\cal L$ indicates the Lie derivative,  and
	\begin{equation}
	\mbox{\boldmath $l$\unboldmath} =\frac{\partial}{\partial v}
	\mbox{\hspace{.5in}}
	\mbox{\boldmath $n$\unboldmath} = \frac{\partial}{\partial u} - s^a
	\frac{\partial}{\partial x^a} \label{3.8}
	\end{equation}
are null vectors.  The notation and the equations are given in
appendix~\ref{appendixa},
more detail may be found in Hayward~\cite{Hayward}.

Our aim is to calculate the Weyl curvature scalars on the intersecting null
hypersurfaces $S^+ $ and $S^-$ by making an ansatz for the free gravitational
data ($k_{ab}$) on these surfaces.  $S^-$ is taken to coincide with the CH at
$v=0$,  while $S^+ $ is an outgoing null hypersurface, parallel to the event
horizon,  crossing $S^-$ near to $P$ in Fig.~\ref{figure2}.  The choice of the
data is
motivated by our understanding of the spherical situation and by the non-linear
perturbation analysis of Ori~\cite{Ori:92}.

\subsubsection*{Assumptions}

The focussing equation on $S^+ $ is [Eq. (\ref{E1})]
	\begin{equation}
	\frac{\partial \theta}{\partial v} = -\frac{1}{2} \theta^2 - \nu \theta
	- \frac{1}{4} \sigma_{ab}\sigma^{ab} \; . \label{3.9}
	\end{equation}
Hayward noticed that making use of coordinate freedom on $S^+ $ (rescaling
$v$),  this equation may be linearised.  Thus choosing $v$ such that
	\begin{equation}
	\nu^+ = -\frac{1}{2} \theta^+ \label{3.10}
	\end{equation}
it is possible to obtain the solution to the Einstein equations on
$S^+ $.  Each of the equations (\ref{E1})-(\ref{E9}) can be integrated (in the
order they appear) provided one knows $k_{ab}$ on this hypersurface, since
	\begin{equation}
	\sigma_{ab}^{+}\sigma^{ab}_{+} = (k_+^{ab} \partial_v k^+_{bc})(k^{cd}_+
	\partial_v
		k^+_{da} ) \; . \label{3.11}
	\end{equation}

We cannot evolve generic initial data from the event horizon of a
non-spherical, rotating black hole so we make an ansatz for this conformal
2-metric. It is inferred from the work of Ori~\cite{Ori:92} that the conformal
metric may be written as
	\begin{equation}
	k_{ab}^+ = k_{ab}^o (x^a) + K^+_{ab}(\ln|v|, x^a) \label{3.12}
	\end{equation}
where $\det(k_{ab}^o ) = 1$ and asymptotically
	\begin{equation}
	 K^+_{ab} \simeq {\cal F}_{ab}(x^c) (-\ln|v|)^{-n}  + \ldots \; .
	\label{3.13}
	\end{equation}
as $v \rightarrow 0$. The functions ${\cal F}_{ab}(x^c)$ are well behaved for
all
$x^a$,
and $n$ is an integer. (The value implied by Ori's work is $n \ge 6$.)
This form is based on the observation that
non-linear metric perturbations decay according to an inverse power-law of
advanced time~\cite{Ori:92},  and the expectation that Hayward's coordinate is
related to external advanced time by
	\begin{equation}
	v \simeq - e^{-\kappa_o w}(1 + \ldots) \label{3.14}
	\end{equation}
as $v \rightarrow 0$ (or $w \rightarrow \infty$).  While this cannot be proven
to
hold,  it seems reasonable as long as the surface $S^+$ does not encounter
a caustic ($\Omega = 0$) at or before the CH -- this condition is satisfied
{\em a postiori} implying, at least, a self-consistent treatment.

\begin{figure}[t]
\leavevmode
\hbox{\epsfxsize=8cm \epsfysize=8cm {\epsffile{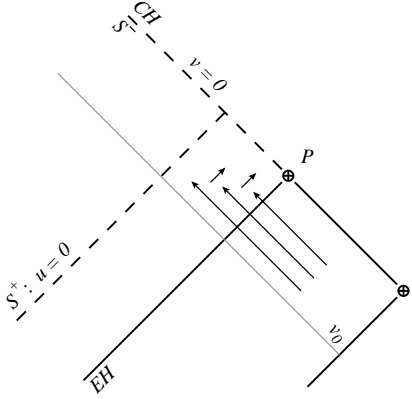}}}
\caption{\label{figure2}A schematic representation of the generic black hole
interior.
The null surface $S^-$ coincides with the Cauchy horizon, CH.  The other null
surface
$S^+$ is at $u=0$ and intersects the Cauchy horizon at $S$ before a caustic
forms.
Initial values of the fields are specified on the intersection between the
hypersurface $v=v_0$ and $S^+$.  Also indicated is the ingoing gravitational
wave tail,
and its scattered component}
\end{figure}

As in the spherical analysis,  we give no detail of the gravitational data on
$S^-$ save to point out that the gravitational energy flux crossing the CH
should not diverge.  In fact, on physical grounds, it is expected that any
radiation from the star will be exponentially redshifted near $P$ in
Fig.~\ref{figure2}.
Furthemore once the incoming radiation gets scattered by the gravitational
potential inside the black hole,  it should only produce a slow contraction of
the CH~\cite{Drozetal}

To summarise,  we assume that a portion of the CH exists which is caustic free
near to $P$ in Fig.~\ref{figure2}.  Furthermore, the gravitational
perturbations
propagating into the hole decay according to an inverse power law of advanced
time which is related to $v$ by (\ref{3.14}).

\subsubsection*{Solution on $S^+$}

With this aspect in hand we can proceed to the solutions of
Eqs.~(\ref{E1})-(\ref{E8}).
Integrating (\ref{3.9})
	\begin{equation}
	\theta^+ = \theta^o(x^a) - \frac{1}{4} \int_{v_0}^v
	\!\! d\tilde{v} \;
	\sigma_{ab}^+\sigma^{ab}_+
	\label{3.15}
	\end{equation}
A (sub) super-script $o$ indicates the initial value of the function at $v_0$
in
$S^+$ (see
Fig. \ref{figure2}).
On this two surface we expect all the
fields to be well behaved,  bounded (and in some cases non-zero) functions of
$x^a$.

	Based on the assumption (\ref{3.12}) and (\ref{3.13})
	 about $k_{ab}^+$ we
can estimate the behaviour of this integral near to the CH.  Firstly note that
the inverse of the conformal metric can be written
	\begin{equation}
	k^{ab}_+ = k^{ab}_o + \epsilon^{ac} \epsilon^{bd} K^+_{cd}
	\end{equation}
where $\epsilon^{bc}$ is the 2-dimensional, antisymmetric matrix with
$\epsilon^{12} =1$.  The functional form of $K^+_{ab}$ implies that
	\begin{equation}
	\partial_v K^+_{ab} = \frac{1}{v} \frac{\partial K^+_{ab}}{\partial \ln
	|v|}
	\end{equation}
will diverge since
	\begin{equation}
	 \frac{\partial K^+_{ab}}{\partial \ln |v|} \simeq n {\cal F}_{ab}
		(-\ln|v|)^{-n-1}
	\end{equation}
as $v\rightarrow 0$.  Since $\sigma_{ab}\sigma^{ab}$ is $v^{-2}$ times a slowly
varying function (near the CH), $\theta^+$ is
asymptotically
	\begin{equation}
	\theta^+ \simeq \theta^0(x^a) +  \frac{n^2 k_o^{ab}
	 k_o^{cd} {\cal F}_{bc} {\cal F}_{da}}{4 v (-\ln|v|)^{2(n+1)}} +
	\ldots
	\end{equation}
which diverges at the CH for the same reason that $\partial_v K^+_{ab}$ does.

It is now straightforward to formally integrate the remaining equations,
however
our main objective is to show which quantities diverge and which remain bounded
at the CH.  Observing that the integral
	\begin{equation}
	I = \int^v \tilde v^{-1} (-\ln|\tilde v|)^{-n} d\tilde v =
	\frac{1}{n-1}\; (- \ln |v|)^{-n+1}
	\label{log}
	\end{equation}
is finite as $v\rightarrow 0$ (provided $n>1$),  it is clear that
	\begin{equation}
	\int^v \!\! d\tilde v \; \left( \tilde v^{-1} K^+_{ab}\right)
	\rightarrow
	\frac{1}{n-1}\; {\cal F}_{ab}(- \ln |v|)^{-n+1}
	\end{equation}
as $v\rightarrow 0$, and remains bounded at the CH.  Each of the equations must
be
treated in turn,  examining the divergent part of the source and its integral
near the CH.  We quickly see that $\Omega$ is bounded at the CH,  and so all
the intrinsic quantities on the two dimensional surfaces of foliation remain
finite all the way to the CH.  To see this,  first integrate (\ref{E2}) to get
	\begin{equation}
	\lambda_+ = \lambda_o - \frac{1}{2} \int_{v_o}^v\!\! d\tilde v  \;
	 \theta^+ \; ,
	\end{equation}
then substituting in the asymptotic form for $\theta^+$ we find
	\begin{eqnarray}
	\int^v\!\! d\tilde v\, \theta^+  &\simeq&   	\frac{n^2}{4} k_o^{ab}
	 k_o^{cd} {\cal F}_{bc} {\cal F}_{da} \int^v\!\! d\tilde v \;
	\tilde v^{-1} (-\ln|\tilde v|)^{-2(n+1)}\nonumber\\
	&\simeq& \frac{n^2 }{4(2n+1)} k_o^{ab}
	 k_o^{cd} {\cal F}_{bc} {\cal F}_{da}(-\ln|v|)^{-(2n+1)}
	\end{eqnarray}
and hence $\lambda_+ \rightarrow$~constant as $v\rightarrow 0$.  The gauge
choice (\ref{3.10})
implies that
	\begin{equation}
	\Omega^+ = \Omega_o e^{-2(\lambda_+ - \lambda_o)}
	\end{equation}
and so $\Omega$ and $k_{ab}$ are both bounded at the CH.  Now continuing the
integration
	\begin{eqnarray}
 	\omega^+_b  &=& \omega^{o}_b e^{2(\lambda_+ - \lambda_{o})}
	\label{3.25}\\
 	&&+ \mbox{$\frac{1}{2}$} { e^{2\lambda_+} }
	\int^{v}_{v_0}\! \! d\tilde v \; e^{-2\lambda_+} (
	\triangle^{a}\sigma_{ab}^+
         -  \mbox{${ 3   \over   2 }$}
         \triangle_b \theta^+  -  \theta^+ \triangle_b
	\lambda_+ ) \; ,
	\nonumber	\\
  	s^b_+   &=&  s_{o}^b + 2 \int^{v}_{v_0}\!\!d\tilde v\; e^{-\lambda_+}
	\omega^b_+ \label{3.26}
	\end{eqnarray}
and
       \begin{eqnarray}
	\widetilde{\theta}^{+} &=& \widetilde{\theta}_{o}
	e^{2(\lambda_{+} - \lambda_o)}
	\\
 	&&+e^{2 \lambda_{+}} \int^{v}_{v_{o}}\! \! d\tilde v \; e^{-3
	\lambda_{+}}
	\biggl( -\mbox{$\frac{1}{2}$} {}^{(2)}\!R + \omega^{a}_{+}
	\omega^{+}_{a} -
	\mbox{$\frac{1}{2}$}
	\triangle^{a}\triangle_{a}\lambda_{+} \nonumber \\
	&&  +  \mbox{$\frac{1}{4}$} \triangle^{a}
	\lambda_{+}\triangle_{a}\lambda_{+} + \omega^{a}_+
	\triangle_{a}\lambda_{+} - \triangle_{a}\omega^{a}_{+}\biggr) \;
	,\nonumber
	\end{eqnarray}
from which it is clear that only finite terms appear in the equation for
$\widetilde{\theta}$,  which therefore cannot diverge on CH.
	\begin{equation}
	\widetilde{\sigma}_{ab}^+ \simeq (\mbox{\rm finite}) - \mbox{$
	\frac{1}{2} $} {e^{\lambda_+}h_{ac}^+ }
	\int^v_{v_0}\!\! d\tilde v\; e^{-\lambda_+}\widetilde{\theta}
	h^{cd} \sigma_{db} \; ,
	\end{equation}
the shear of the ingoing null rays orthogonal to $S(u=0,v)$,  is also bounded
at the CH. Finally
	\begin{eqnarray}
	\widetilde{\nu}_+ &=& \widetilde{\nu}_{o} + \int^{v}_{v_{o}}\! \!
	d\tilde
	v \;
	\left\{ \mbox{$\frac{1}{4}$}
	\sigma^{+}_{ab}\widetilde{\sigma}^{ab}_{+} -\mbox{$\frac{1}{2}$}
	\theta^{+}\widetilde{\theta}^{+} \right. \\
	&& \mbox{\hspace{-0.2in}} + \left. e^{- \lambda_{+}}
	(- \mbox{$\frac{1}{2}$} {}^{(2)}R + 3\omega_{a}^{+} \omega^{a}_{+} -
	\mbox{$\frac{1}{4}$}  \triangle_{a} \lambda_{+}
	\triangle^{a} \lambda_{+} - \omega^{a}_{+} \triangle_{a}\lambda_{+})
	\right\} \nonumber
	\end{eqnarray}
which is finite despite the presence of divergent terms in the integrand [which
look
like $v^{-1} \times $ inverse powers of $ (-\ln|v|)$, see Eq.
(\ref{log})].
  This analysis implies that $\theta^+$ and $\sigma^+_{ab}$ diverge
inside the black hole,  and the divergence is of the same nature as in the
spherical model  once $\sigma_{ab} \sigma^{ab}$ is interpreted as the
gravitational energy crossing $S^+$.

\section{The Cauchy horizon singularity} \label{section5}

The solution obtained in the previous section can be used to show that a scalar
curvature singularity is present on the CH due to the blueshifted influx of
radiation.  On $S^-$ the shift vector $s^a$ is set to zero by a coordinate
transformation $\tilde x^a = \tilde x^a (u,x^b)$ without altering the form of
the metric (\ref{3.2}).  (These coordinates exist provided $s^a$ is regular on
$S^-$ in the original coordinate system.  Eq. (\ref{3.26}) suggests that this
is
true.)  It is assumed that this transformation was made in
section~\ref{section2}; this in no way alters the previous analysis.

Hayward's scheme to integrate the equations on $S^-$ can now be completed.
Making the gauge choice
	\begin{equation}
	\widetilde{\nu}^- = - \mbox{$1 \over 2$} \widetilde{\theta}^-\; ,
	\end{equation}
Eq.~(\ref{E11}) implies that
	\begin{equation}
	\widetilde{\theta}^- = \left. \widetilde{\theta}^+\right|_{v=0} -
	\mbox{$1 \over 4$} \int_{0}^u \!\! d\tilde u \;
		(k_-^{ab} \partial_{\tilde u} k^-_{bc})(k^{cd}_-
	\partial_{\tilde u}
		k^-_{da} )	 \; .
	\end{equation}
As mentioned in the introduction $\widetilde{\theta}^-$ (and all other
quantities) are
formally known once $k_{ab}^-(u,x^c)$ is specified.  Our analysis assumes that
$k_{ab}^-$ is a slowly varying function of $u$ which tends to a non-degenerate
value as $u\rightarrow - \infty$; that is,  near to $P$ (in Fig.~\ref{figure2})
$k_{ab}$ should approach the value obtained on the CH in Kerr spacetime.
Furthermore $\widetilde{\theta}^- \rightarrow 0$ and the spacetime should be
asymptotically Kerr in the limit $u\rightarrow - \infty$.   It is not a generic
situation to have  $\widetilde{\theta}^- =0$ and/or $\widetilde{\sigma}_{ab}^-
=
0$ on the CH,
since some radiation (both from the star,  and from backscattering of the
gravitational wave tail inside the black hole\cite{Ori:92,Drozetal})  will
usually be present.  This discussion implies that there should exist a caustic
free portion of the CH near to $P$ on which our analysis is valid.

The integration now proceeds as it did for $S^+$,  and in particular we arrive
at expressions for $\theta^-$ and $\sigma_{ab}^-$
        \begin{eqnarray}
        \theta^- &=& e^{2 \lambda_-} \left[e^{-2 \lambda_{+}}
	\theta^{+} \right]_{v=0}  \\
	&&+
         e^{2 \lambda_-}\int^{u}_{0}\!\! d\tilde u\;  e^{-3 \lambda}\bigl(-
	\mbox{$\frac{1}{2}$} ^{(2)}R    +  \omega_a \omega^a \nonumber \\
	&&
        \mbox{\hspace{-0.1in}} - \mbox{$\frac{1}{2}$}
	\triangle^a\triangle_a \lambda   \mbox{\ } +
        \mbox{$\frac{1}{4}$} \triangle_a\lambda \triangle^a\lambda
         - \omega^a \triangle_a
        \lambda
        + \triangle_a \omega^{a} \bigr)  \;\nonumber
        \end{eqnarray}
and
        \begin{eqnarray}
        \sigma_{ab}^- &=& e^{\lambda_-} h_{fa}^- \left[ e^{-\lambda_+} h^{fd}_+
                        \sigma_{db}^+\right]_{v=0} \\
        && \mbox{\hspace{-0.1in}}- \mbox{$\frac{1}{2}$}
        h_{fa}^- e^{\lambda_-} \int^u_0\!\! d\tilde u \;
	\bigl( e^{-\lambda} \theta h^{df}
	\widetilde{\sigma}_{db}
        -4 e^{-2 \lambda}  h^{df} \bigl\{
	\omega_{(d}\,\omega_{b)}\nonumber\\
         && \mbox{\hspace{-0.1in}}- \mbox{$\frac{1}{2}$}
        \triangle_{(d} \triangle_{b)} \lambda   +
        \mbox{$\frac{1}{4}$} \triangle_{(d} \lambda  \triangle_{b)}
   	\lambda -
        \omega_{(d} \triangle_{b)} \lambda  + \triangle_{(d}
	\omega_{b)} \bigr\}\bigr)
        \nonumber\\ &&
        \mbox{\hspace{-0.1in}} -
        h_{ab}^- e^{\lambda_-} \int_0^u\!\! d\tilde u\; e^{-2\lambda}\bigl(
	\omega^{c}\omega_{c} -
        \mbox{$\frac{1}{2}$} \triangle_c\triangle^c \lambda +
	\mbox{$\frac{1}{4}$}
        \triangle_c\lambda\triangle^c\lambda \nonumber \\
        && \mbox{\hspace{-0.1in}}  -  \omega^{c} \triangle_c\lambda
	+
        \triangle_c \omega^{c}\bigr)  \; . \nonumber
        \end{eqnarray}

 We saw in the previous section how
	\begin{equation}
	\lim_{v\rightarrow 0} \left\{
	\begin{array}{l}
	\sigma^+_{ab} \\
	\theta^+
	\end{array} \right\}
	\rightarrow \infty \; ,
	\end{equation}
thus $\theta^-$ and
$\sigma_{ab}^-$ are generically unbounded on the CH. Of the remaining fields,
only
$\nu$ contains divergent quantities. From Eq.~(\ref{E19})
we see that
	\begin{eqnarray}
	\nu_- \simeq \int^{u}_{0} \!\!d \tilde u\; ({1 \over 4} \sigma^{-}_{ab}
	\widetilde{\sigma}_{-}^{ab}
	-{1 \over 2} \theta^- \widetilde{\theta}_-) \, + \, \ldots
	\;
	,
	\end{eqnarray}
which is actually infinite on CH.
Imposing the field equations on the Weyl scalars one finds that all but
$\Psi_4$
contain divergent quantities,  and hence the CH is singular.

It is possible to say a little more about the nature of the singularity by
examining the rate of growth of the curvature along $S^+$.  Once again,
imposing
the field equations in Eq.~(\ref{psi0}) gives
	\begin{equation}
	\Psi_0^+ = \mbox{$1\over 4$} e^{2\lambda} m^am^b \left.
	\left\{ 2 {\cal L}_{l} \sigma_{ab}
	+ 2\nu \sigma_{ab} - \sigma_{am}h^{mn}\sigma_{nb}\right\} \right|_+\; .
	\end{equation}
Using $\sigma_{ab}^+$, etc.\@ the asymptotic expression is
	\begin{equation}
	\Psi_0^+ \simeq (\mbox{finite})\times \left\{ v^2 (-\ln |v|)^{n+1}
	\right\}^{-1}
	\label{4.7}
	\end{equation}
as $v\rightarrow 0$. This is the contribution from ${\cal L}_{l}
\sigma_{ab}^+$,
which is
damped by the smallest number of powers of $(-\ln|v|)$.  In a similar manner
the
leading behaviour can be extracted from $\Psi_1$ and $\Psi_2$ giving
	\begin{eqnarray}
	\Psi_1^+ &\simeq& \mbox{(finite})\times \left\{ v\, (-\ln |v|)^{n+1}
	\right\}^{-1}\\
	\Psi_2^+ &\simeq& \mbox{(finite})\times \left\{ v\, (-\ln |v|)^{n+1}
	\right\}^{-1}
	\label{4.9}
	\end{eqnarray}
where these leading contributions arise due to the presence of terms involving
the shear ($\sigma_{ab}^+$)  of the surface $S^+$.  A little more work shows
that $\Psi_3^+$ is finite on the CH;  notice that eq.~(\ref{3.26}) combined
with
the choice $s^a \rightarrow 0$ on $S^-$,  and knowledge that $\lambda^+$ and
$\omega_a^+$ are finite on $S^+\cap S^-$ implies that 	$s^+_a \sim v$ as $v
\rightarrow 0$,  hence $s^m_+ \Sigma_{mn}^+s^n_+ \rightarrow 0$ on $S^-$.
Finally $\Psi_4$ is independent of $\theta^+$ and $\sigma_{ab}^+$ and therefore
is finite.
These results imply that the square of the Weyl tensor is dominated by
$\Psi_0$,
	\begin{eqnarray}
	C_{\alpha\beta\gamma\delta}C^{\alpha\beta\gamma\delta}
	&\simeq& 8 \,(\Psi_0\, \Psi_4 + \overline{\Psi}_0\,\overline{\Psi}_4)
	 + \ldots \nonumber \\
	&\simeq& (\mbox{finite})\times \left\{ v^2 (-\ln |v|)^{n+1}
	\right\}^{-1}
	  + \ldots \label{Weylsqr}
	\end{eqnarray}
as $v\rightarrow 0$.

\section{Discussion}\label{section6}

The treatment of the non-spherical gravitational collapse is generally very
difficult,  however it seems likely that the external field settles down to
that
of a Kerr black hole.  Inside the black hole some progress may also be made by
advancing two assumptions: 1)  There exists a caustic free portion of the CH,
which is a null hypersurface. 2)  There is an ingoing gravitational wave tail
which gets blueshifted by an infinite amount at the CH (this is achieved here
by
our ansatz for $k_{ab}^+$).  Indeed using Hayward's formalism the nature of the
CH can be investigated in some detail.

Following from the analysis in sections~\ref{section2} and~\ref{section5} we
conclude that the CH is the locus of a scalar curvature singularity,  and the
curvature is an integrable function of the advanced time.
Ori~\cite{Ori:91,Ori:92} has
argued that the latter property of the curvature suggests
an
extension of the spacetime through the singularity may be
possible.  It is worth commenting a little further on
this point. We have demonstrated the asymptotic values of
each of the Weyl scalars,  in particular it has been shown that $\Psi_0$
contains
the leading divergences.  The algebraic classification of the Weyl curvature is
obtained by solving the quartic
	\begin{equation}
	\Psi_0 a^4 + \Psi_1 a^3 + \Psi_2 a^2 + \Psi_3 a + \Psi_4 = 0
	\end{equation}
for $a$ and examining the degeneracy of its roots~\cite{Chadra:bh}.  Based on
eqs.~(\ref{4.7})-(\ref{4.9}) there is a four-fold degeneracy as $v\rightarrow
0$
implying that the gravitational field is asymptotically type N,  with repeated
principal
null direction $ \mbox{\boldmath
$k$\unboldmath} = \mbox{\boldmath
$n$\unboldmath} + a \mbox{\boldmath
$m$\unboldmath} + a^*\bar{\mbox{\boldmath
$m$\unboldmath}} + a a^* \mbox{\boldmath
$l$\unboldmath}$ (where $a \rightarrow 0$ as $v\rightarrow 0$).  It is worth
noting that gravitational shock waves are of this
algebraic
type (and are the only curvatures which may be confined to a thin skin without
a surface layer of matter being present~\cite{Israel});  therefore the
intuitive picture of the singularity which emerges is that of a gravitational
shock propagating along the CH.   These results might be taken to support
arguments for the existence of a classical continuation of the geometry beyond
the CH singularity. However, this is a highly speculative point and it should
be noticed that the Weyl curvature contains contributions which are not present
in an exactly type~N spacetime.   Indeed these contributions diverge at the CH
and   are manifest in the square of the Weyl tensor~(\ref{Weylsqr}) (which
vanishes for a pure gravitational shock).  Thus a classical continuation is
unlikely, as pointed out in~\cite{BBIP},  since non-classical matter is needed
to confine the divergent curvature to a thin layer along the CH.

To conclude,  the {\em hairy} singularity inside a generic black hole is
dominated by the propagating gravitational wave tail of the collapse (which
decays in the exterior of the black hole).

\acknowledgements
It is a pleasure to thank Werner Israel for comments and discussions on the
interpretation of our results.
CMC is funded by the EPSRC of Great Britain, and gratefully
acknowledges the hospitality of
Ian Pinkstone and James Briginshaw of DAMTP, Cambridge where
part of this work was performed. PRB is grateful to Peter Hogan and
Adrian
Ottewill for their kind hospitality at University College, Dublin where part of
this work
was carried out.

\onecolumn
\widetext


\appendix


\section{Notation}\label{appendixa}

In this appendix we summarise the dual null formalism of
Hayward~\cite{Hayward}.
The reader is referred to his articles for more technical details.
This approach,  based on a 2+2 decomposition,  assumes that the spacetime can
be foliated locally by compact orientable 2-surfaces S.  Therefore given a
Lorentzian manifold $\cal M$ (with signature $-+++$) we assume that there
exists
a smooth embedding
$\phi : S \times [0,U) \times [0,V) \rightarrow {\cal M}$ with the induced
metric,
$h_{ab}$, on  $S$ being spatial.   Latin indices range over $\{1,2\}$.  In
terms
of a basis which is Lie-propagated along $\partial/\partial u$ and
$\partial/\partial v$ the spacetime line element may be written as in
Eq.~(\ref{3.2}).
The Lie derivative along a vector $\xi$ is denoted $ {\cal L}_{\xi} $.  The
natural
covariant derivative of the metric $h_{ab}$ is $\triangle_a $, and the Ricci
scalar
is denoted by $ {{^{(2)}}R} $. Tensor valued quantities on the 2-surface $S$
are
denoted $X{^{a \dots}}_{b \dots}$ (sometimes we will include a superscript
$(2)$
to emphasise the 2-dimensional covariant nature).  Introducing the null vectors
	\begin{equation}
	l^{\mu} = (1,0,0,0), ~~~~
	n^{\mu} = (0,1, -s^{a}) \; .
	\end{equation}
we define
	\begin{equation}
	\Sigma_{ab} = {\cal L}_{l} h_{ab}\; , \mbox{\hspace{0.3in}}
	\widetilde{\Sigma}_{ab} = {\cal L}_{n} h_{ab} \; .
	\end{equation}


To write the Einstein field equations in a first order form Hayward introduces
the extrinsic fields:
	\begin{eqnarray}
		\theta  &=& \hbox to .75in{$\displaystyle
			\mbox{$1 \over 2$} h^{ab} \Sigma_{ab}$,\hfill}
	\mbox{\hspace{.5in}}
		\hbox to .19in{\hfill$\displaystyle {\widetilde{\theta}}$}
		=  \mbox{$1 \over 2$} h^{ab} \widetilde{\Sigma}_{ab} , \\
		\sigma_{ab}  &=&  \hbox to .75in{$\displaystyle \Sigma_{ab} -
				\theta h_{ab}$,\hfill} \mbox{\hspace{.5in}}
		\hbox to .19in{\hfill$\displaystyle \widetilde{\sigma}_{ab}$}
		 = \widetilde{\Sigma}_{ab} - \widetilde{\theta} h_{ab} ,  \\
		\nu  &=&  \hbox to .75in{$\displaystyle {\cal L}_{l}
	\lambda$,\hfill}
	\mbox{\hspace{.5in}}
		\hbox to .19in{\hfill$\displaystyle\widetilde{\nu} $}  = {\cal
	L}_{n}  \lambda \; ,\\
		\omega_a  &=&  \mbox{$1 \over 2$} e^{\lambda} h_{ab} {\cal
	L}_{l} s^b \; .
	\end{eqnarray}
These quantities are readily interpreted geometrically; $ (\theta,
\widetilde{\theta}) $
are
the expansions in the null directions normal to the foliation  $ { S} $.
The tensors $ (\sigma_{ab}, \widetilde{\sigma}_{ab}) $ are the shears, which
are
traceless
by definition,  the vector $ \omega_a $  is the twist.
The two functions $ (\nu, \widetilde{\nu}) $ are called the inaffinities and
measure the non-affine quality of the coordinates $u$ and $v$.
Then the Einstein equations and contracted Bianchi identities are: the
\boldmath
$l$ \unboldmath
equations:
	\begin{eqnarray}
          {\cal L}_{l} \hbox to .19in{$\displaystyle \theta$\hfill} \,
	    =&& -{ \mbox{$1 \over 2$} } \theta^2
            -    \nu \theta
            -    {\mbox{$1 \over 4$}} \sigma_{ab}  \sigma^{ab} \label{E1} \\
  	  {\cal L}_{l} \hbox to .19in{$\displaystyle  \lambda $\hfill}\,   =&&
	\nu
	\label{E2} \\
	  {\cal L}_{l} \hbox to .19in{$\displaystyle  \Omega $\hfill}\,\, =&&
	\theta\, \Omega
	\label{E3} \\
	  {\cal L}_{l} \hbox to .19in{$\displaystyle  k_{ab} $\hfill}\, =&&
	\Omega^{-1}
	\sigma_{ab} \label{E4} \\
  	  {\cal L}_{l} \hbox to .19in{$\displaystyle  \omega_a $\hfill}\,  =&&
	-\theta
		\, \omega_a
           +     {\mbox{$1 \over 2$}} \triangle^{b} \sigma_{ba}
           +     {\mbox{$1 \over 2$}}\triangle_a (\nu - \theta)
           -     {\mbox{$1 \over 2$}} \theta \triangle_ a\lambda \label{E5} \\
	  {\cal L}_{l} \hbox to .19in{$\displaystyle  s^a$}\,  =&&  2
	e^{-\lambda} \omega^{a}
	\label{E6} \\
	 {\cal L}_{l} \hbox to .19in{$\displaystyle
	\widetilde{\theta}$\hfill}\,
		=&&  -\theta \widetilde{\theta}%
                 +     e^{- \lambda}(- { \mbox{$1 \over 2$}}{ {^{(2)}}R}
                 +     \omega_a \omega^{a}
                 -     {\mbox{$1 \over 2$}} \triangle^a\triangle_a \lambda
                 +     { \mbox{$1 \over 4$}}\triangle_a \lambda\,
	\triangle^{a} \lambda
                 +     \omega^{a}\, \triangle_a \lambda
                 -     \triangle_a \omega^{a} )  \label{E7} \\
	 {\cal L}_{l} \hbox to .19in{$\displaystyle
	\widetilde{\sigma}_{ab}$\hfill}\,
		   =&&  \sigma_{ac}\, h^{cd}\, \widetilde{\sigma}_{db}
                   +    {\mbox{$1 \over 2$}} \theta \, \widetilde{\sigma}_{ab}
                   -    {\mbox{$1 \over 2$}} \widetilde{\theta}\,  \sigma_{ab}
	\nonumber\\
                   && \mbox{\ \ \ } +  2 e^{- \lambda}( \omega_a \omega_b
                   -    {\mbox{$1 \over 2$}}\triangle_{(a } \triangle_{b)}
	\lambda
                   +    {\mbox{$1 \over 4$}} \triangle_{(a} \lambda
	\triangle_{b)}
		   \lambda
                   +    \omega_{(a} \triangle_{b)} \lambda
                   -    \triangle_{(a}  \omega_{b)}) \nonumber\\
                   && \mbox{\ \ \ } -  e^{- \lambda} \, h_{ab}\, (\omega_c
	\omega^{c}
                   -    {\mbox{$1 \over 2$}} \triangle^{c}\triangle_{c} \lambda
                   +    { \mbox{$1 \over 4$}} \triangle_c \lambda\,
	\triangle^{c}\lambda
                   +    \omega^{c} \triangle_c\lambda
                   -     \triangle_c \omega^{c})  \label{E8} \\
	 {\cal L}_{l} \hbox to .19in{$\displaystyle  \widetilde{\nu}$\hfill}\,
		=&&  {\mbox{$1 \over 4$}} \sigma_{ab} \widetilde{\sigma}^{ab}
                -    { \mbox{$1 \over 2$}} \theta \widetilde{\theta}
                +    e^{ - \lambda} ({ -\mbox{$1 \over 2$}}{{^{(2)}}R}
                +    3 \omega_a \omega^{a}
                -    { \mbox{$1 \over 4$}}\triangle_a \lambda\,
	\triangle^{a}\lambda
                 -   \omega^{a} \triangle_a \lambda )  \label{E9}
	\end{eqnarray}
and the \boldmath $n$ \unboldmath equations:
	\begin{eqnarray}
          {\cal L}_{n} \hbox to .19in{$\displaystyle \widetilde{\theta}$\hfill}
	\,  =&& -{
	\mbox{$1 \over 2$} } \widetilde{\theta}^2
            -    \widetilde{\nu} \widetilde{\theta}
            -    {\mbox{$1 \over 4$}} \widetilde{\sigma}_{ab}
	\widetilde{\sigma}^{ab}
	\label{E11} \\
  	  {\cal L}_{n} \hbox to .19in{$\displaystyle \lambda$\hfill}  =&&
	\widetilde{\nu}  \label{E12}
	\\
	  {\cal L}_{n} \hbox to .19in{$\displaystyle \Omega$\hfill} =&&
	\widetilde{\theta} \, \Omega
	\label{E13} \\
	  {\cal L}_{n} \hbox to .19in{$\displaystyle k_{ab}$\hfill} =&&
	\Omega^{-1}
	\widetilde{\sigma}_{ab} \label{E14} \\
  	  {\cal L}_{n} \hbox to .19in{$\displaystyle \omega_a$\hfill}  =&&
	-\widetilde{\theta}
	\, \omega_a
           -     {\mbox{$1 \over 2$}} \triangle^{b} \widetilde{\sigma}_{ba}
           -     {\mbox{$1 \over 2$}}\triangle_a (\widetilde{\nu} -
	\widetilde{\theta})
           +     {\mbox{$1 \over 2$}} \widetilde{\theta} \triangle_ a\lambda
	\label{E15}
	\\
	 {\cal L}_{n} \hbox to .19in{$\displaystyle \theta$\hfill} =&&  -\theta
	\widetilde{\theta}
                 +     e^{- \lambda}(- { \mbox{$1 \over 2$}}{ {^{(2)}}R}
                 +     \omega_a \omega^{a}
                 -     {\mbox{$1 \over 2$}} \triangle^a\triangle_a \lambda
                 +     { \mbox{$1 \over 4$}}\triangle_a \lambda\,
	\triangle^{a} \lambda
                 -     \omega^{a}\, \triangle_a \lambda
                 +     \triangle_a \omega^{a} )  \label{E17} \\
	 {\cal L}_{n} \hbox to .19in{$\displaystyle \sigma_{ab}$\hfill}
		=&&  \widetilde{\sigma}_{ac}\,  h^{cd} \, \sigma_{db}
                   +    {\mbox{$1 \over 2$}} \widetilde{\theta}\,  \sigma_{ab}
                   -    {\mbox{$1 \over 2$}} \theta\, \widetilde{\sigma}_{ab}
	\nonumber\\
                   && \mbox{\ \ \ } +  2 e^{- \lambda}( \omega_a \omega_b
                   -    {\mbox{$1 \over 2$}}\triangle_{(a } \triangle_{b)}
	\lambda
                   +    {\mbox{$1 \over 4$}} \triangle_{(a} \lambda
	\triangle_{b)}
		   \lambda
                   -    \omega_{(a} \triangle_{b)} \lambda
                   +    \triangle_{(a}  \omega_{b)}) \nonumber\\
                   &&  \mbox{\ \ \ } - e^{- \lambda} \, h_{ab}\, (\omega_c
	\omega^{c}
                   -    {\mbox{$1 \over 2$}} \triangle^{c}\triangle_{c} \lambda
                   +    { \mbox{$1 \over 4$}} \triangle_c \lambda\,
	\triangle^{c}\lambda
                   -    \omega^{c} \triangle_c\lambda
                   +     \triangle_c \omega^{c})  \label{E18} \\
	 {\cal L}_{n} \hbox to .19in{$\displaystyle \nu$\hfill}
		=&&  {\mbox{$1 \over 4$}} \sigma_{ab} \widetilde{\sigma}^{ab}
                -    { \mbox{$1 \over 2$}} \theta \widetilde{\theta}
                +    e^{ - \lambda} ({ -\mbox{$1 \over 2$}}{{^{(2)}}R}
                +    3 \omega_a \omega^{a}
                -    { \mbox{$1 \over 4$}}\triangle_a \lambda\,
	\triangle^{a}\lambda
                 +   \omega^{a} \triangle_a \lambda )  \label{E19}
	\end{eqnarray}


\section {Riemann Tensor And Weyl Scalars}\label{appendixc}

\subsection{Riemann Tensor}\label{riemann}
Here we list all the components of the Riemann tensor for the metric
(\ref{3.2}):
	\begin{eqnarray}
	R^{v}_{vuv} =&&-          {\cal L}_{l}\, { \widetilde{\nu} }
               -        s^{a}  \triangle_a \nu
               +        { \mbox{$1 \over 2$} }  e^{\lambda}  s^{a} \Sigma_{ab}
		s^{b} \nu
               +        2 s^{a} {\cal L}_{l}\, \omega_a
               -      e^{- \lambda} \omega_a \triangle^{a} \lambda
               +        3 e^{ - \lambda} \omega^{a}  \omega_a
               +        s^{a}  \Sigma_{ab} \omega^{a} \nonumber\\
               &&\mbox{\ \ \ }+      { \mbox{$1 \over 2$} }e^{\lambda} s^{a}
	\left( {\cal L}_{l}\, \Sigma_{ab}\right)
		s^{b}
               +        { \mbox{$1 \over 2$} } s^{a}  \Sigma_{ab}\,
	\triangle^{b}
		\lambda
               -        { \mbox{$1 \over 4$} } e^{ - \lambda } \triangle^{a}
		\lambda \,  \triangle_a \lambda
               -      { \mbox{$1 \over 4$}} e^{\lambda} s^{a} \Sigma_{ab}
		h^{bc} \Sigma_{cd} s^{d} \\
	R^{v}_{vua} = &&  -  {\mbox{$1 \over 2$}} \triangle_{a}\nu
            +        {\mbox{$1 \over 2$}}e^{\lambda} \nu s^{m}\Sigma_{ma}
            +        {\mbox{$1 \over 2$}} \omega^{m}\Sigma_{ma}
            +     {\cal L}_{l}\, \omega_{a}
            +        { \mbox{$1 \over 2$} } e^{\lambda} s^{m}{\cal L}_{l}\,
	\Sigma_{ma}
            -        {\mbox{$1 \over
		4$}}e^{\lambda}\Sigma_{am}h^{mn}\Sigma_{nq}s^{q}
		+     {\mbox{$1 \over 4$}}\triangle^{m} \lambda \, \Sigma_{ma}\\
	R^{v}_{aub} =&&       { \mbox{$1 \over 2$} } e^{\lambda} {\cal L}_{l}\,
	\Sigma_{ab}
               +      { \mbox{$1 \over 2$} } e^{\lambda}\nu \Sigma_{ab}
               -      { \mbox{$1 \over 4$} } e^{\lambda} \Sigma_{am} h^{mn}
		\Sigma_{nb}
		\\
	R^{v}_{vab} =&&  {\mbox{$1 \over 2$}}e^{\lambda} s^{m}\Sigma_{m[b}
	\triangle_{a]}\lambda
            +    e^{\lambda}s^{m} \triangle_{[a}\Sigma_{b]m}
            -  e^{\lambda}s^{m}\Sigma_{m[a}\omega_{b]}
            +    2 \triangle_{[a}\omega_{b]}
            +    {\mbox{$1 \over
	2$}}e^{\lambda}\Sigma_{n[a}\widetilde{\Sigma}_{b]m}h^{nm}
		\\
	R^{c}_{abd} =&&   - e^{\lambda}s^{c}\triangle_{[b}\Sigma_{d]a}
            -       \mbox{$1 \over
		2$}e^{\lambda}s^{c}\Sigma_{a[d}\triangle_{b]}\lambda
		  +    {\mbox{$1 \over 2$}} e^{\lambda} h^{cm}
		(\Sigma_{a[d}\, \widetilde{\Sigma}_{b]m}
            -       \Sigma_{m[d}\, \widetilde{\Sigma}_{b]a})
            -       e^{\lambda}s^{c}\Sigma_{a[d}\omega_{b]}
            +       {}^{(2)}R{^{c}}_{abd} \\
	R^{v}_{acd} =&&   {\mbox{$1 \over
	2$}}e^{\lambda}\triangle_{[c}\lambda\Sigma_{d]a}
            +     e^{\lambda}\triangle_{[c}\Sigma_{d]a}
            +     e^{\lambda}\Sigma_{a[d}\omega_{c]} \\
	R^{v}_{avb} =&&   {\mbox{$1 \over 2$}}e^{\lambda} {\cal
	L}_{n}\,\Sigma_{ab}
            +     {\mbox{$1 \over 2$}}e^{\lambda}s^{m}(\triangle_{m}\Sigma_{ab}
            -     \triangle_{b}\Sigma_{am})
            +   {\mbox{$1 \over 2$}}\triangle_{a}\triangle_{b}\lambda
            -     {\mbox{$1 \over 4$}}\triangle_{a}\lambda\,
	\triangle_{b}\lambda
            -     \triangle_{b}\omega_{a} \nonumber \\
            &&\mbox{\ \ \ }+   {\mbox{$1 \over 2$}}e^{\lambda}\Sigma_{ab}w^{c}
	s_c
            +     {\mbox{$1 \over
	4$}}e^{\lambda}s^{c}\triangle_c\lambda\, \Sigma_{ab}
            -     {\mbox{$1 \over
	4$}}e^{\lambda}\Sigma_{mb}h^{mn}\widetilde{\Sigma}_{na}
            +   {\mbox{$1 \over 2$}}\omega_{a}\triangle_{b}\lambda
            +     {\mbox{$1 \over 2$}}\omega_{b} \triangle_{a}\lambda
            -     \omega_{a}\omega_{b}  \nonumber \\
            &&\mbox{\ \ \ }-   {\mbox{$1 \over
	2$}}e^{\lambda}\omega_{b}s^{m}\Sigma_{am}
            -     {\mbox{$1 \over
	4$}}e^{\lambda}\triangle_{b}\lambda\, \Sigma_{ma}s^{m} \\
	R^{v}_{vva} =&&  {\mbox{$1 \over 2$}}e^{\lambda} s^{m}{\cal
	L}_{n}\,\Sigma_{ma} +
		{\cal L}_{n}\,\omega_{a}
            +    {\mbox{$1 \over
	2$}}e^{\lambda}s^{m}\left( s^{n}\triangle_n\right)\Sigma_{ma}
            -  \mbox{$1 \over
		2$}e^{\lambda}\Sigma_{ma}h^{mn}\widetilde{\Sigma}_{nq}s^{q}
            +      {\mbox{$1 \over 2$}}\triangle_{a}\widetilde{\nu}
            +    {\mbox{$1 \over 2$}}\left(s^{m}\triangle_m\right)
	\triangle_{a}\lambda
	\nonumber
		\\
            &&\mbox{\ \ \ }-  2s^{m}\triangle_{a}\omega_m
            -    {\mbox{$1 \over 4$}}e^{\lambda}\triangle_{a}\lambda\,
		s^{m}\Sigma_{mn}s^{n}
            -    {\mbox{$1 \over 2$}}e^{\lambda}
		s^{m}\left( \triangle_{a}\Sigma_{mn}\right) s^{n}
            -  {\mbox{$1 \over 4$}} s^{m}\triangle_m \lambda\,
		\triangle_{a}\lambda
            +    {\mbox{$1 \over 4$}} e^{\lambda}s^{m}\Sigma_{ma}
		\, s^{n}\triangle_n \lambda
			\nonumber \\
            &&\mbox{\ \ \ }
	+    {\mbox{$1 \over 2$}}s^{m}\triangle_m\lambda\omega_{a}
	-  {\mbox{$1 \over 4$}}\widetilde{\Sigma}_{am}\triangle^{m}\lambda
	        -    {\mbox{$1 \over 2$}}e^{\lambda} \left(s^{m}\Sigma_{mn}
	s^{n}\right) \omega_{a}
            +    {\mbox{$1 \over 4$}} e^{\lambda}s^{m}\Sigma_{mn} h^{nq}
		\widetilde{\Sigma}_{aq}
            +  {\mbox{$1 \over 2$}}\left( s^{m} \omega_m \right)
	\triangle_{a}\lambda
	\nonumber\\
            &&\mbox{\ \ \ }
	    +    {\mbox{$1 \over 2$}}e^{\lambda}\left(s^{n} \omega_n \right)
	s^{m} \Sigma_{ma}
            -   \left( s^{m} \omega_m \right) \omega_{a}
            +    {\mbox{$1 \over 2$}}\omega^{m} \widetilde{\Sigma}_{ma}
	+  s^{m}\triangle_m \omega_{a}\\
	R^{u}_{avb} =&&  {\mbox{$1 \over
	2$}}e^{\lambda}\widetilde{\Sigma}_{ab}\widetilde{\nu}
            +    \mbox{$1 \over
		4$}e^{\lambda}s^{m}\triangle_m\lambda\, \widetilde{\Sigma}_{ab}
            +    {\mbox{$1 \over 2$}}e^{\lambda}{\cal
	L}_{n}\,\widetilde{\Sigma}_{ab}
            +  {\mbox{$1 \over 2$}}e^{\lambda}s^{m}\triangle_m
	\widetilde{\Sigma}_{ab}
            -    {\mbox{$1 \over 4$}} e^{\lambda}\triangle_{b}\lambda\,
		s^{m}\widetilde{\Sigma}_{am}
            -    {\mbox{$1 \over
		2$}}e^{\lambda}s^{m}\, \triangle_{b}\widetilde{\Sigma}_{am}
		\nonumber \\
            &&\mbox{\ \ \ }-  {\mbox{$1 \over 2$}}e^{\lambda}\left( s_m
	w^{m}\right) \widetilde{\Sigma}_{ab}
            -    \mbox{$1 \over
	4$}e^{\lambda}\widetilde{\Sigma}_{bm}h^{mn}\widetilde{\Sigma}_{an}
            +    {\mbox{$1 \over
	2$}}e^{\lambda}s^{m}\widetilde{\Sigma}_{am}\omega_{b}\\
	R^{c}_{avb}=&&   {\mbox{$1 \over
	2$}}h^{cm}(\triangle_{a}\widetilde{\Sigma}_{mb}
		-\triangle_{m}\widetilde{\Sigma}_{ab})
           -     h^{cn}s_{m}{}^{(2)} R{^{m}}_{ban}
           -   {\mbox{$1 \over 2$}}e^{\lambda}s^{c}{\cal L}_{n}\,\Sigma_{ab}
           -     {\mbox{$1 \over
	2$}}e^{\lambda}s^{c}s^{m}\triangle_m\Sigma_{ab}
           -     {\mbox{$1 \over 2$}}s^{c}\triangle_{b}\triangle_{a}\lambda
	\nonumber \\
           &&\mbox{\ \ \ }+   {\mbox{$1 \over 4$}}
	e^{\lambda}s^{c}\triangle_{b}\lambda\,
		s^{m}\Sigma_{am}
           +     {\mbox{$1 \over
	2$}}e^{\lambda}s^{c}s^{m}\triangle_{b}\Sigma_{am}
           +     s^{c}\triangle_{b}\omega_{a}
           +   {\mbox{$1 \over 4$}}s^{c}\triangle_{a}\lambda\,
	\triangle_{b}\lambda
           -     {\mbox{$1 \over 2$}}s^{c} \omega_{b} \triangle_{a}\lambda
           +     {\mbox{$1 \over 4$}}\triangle_{a}\lambda\,  h^{cm}
	\widetilde{\Sigma}_{mb}
		\nonumber \\
           &&\mbox{\ \ \ }+   {\mbox{$1 \over
	2$}}e^{\lambda}s^{m}\Sigma_{am}s^{c}w_{b}
           -     {1 \over
		4}e^{\lambda}s^{n}\Sigma_{an}h^{cm}\widetilde{\Sigma}_{mb}
           -   {\mbox{$1 \over 2$}}s^{c} \omega_{a} \triangle_{b}\lambda
      +     s^{c}\omega_{a}\omega_{b}
           -     {\mbox{$1 \over 2$}}h^{cm}\widetilde{\Sigma}_{mb}\omega_{a}
	\nonumber \\
           &&\mbox{\ \ \ }-   {\mbox{$1 \over
	2$}}e^{\lambda} \omega^{m} s_m\, s^{c}\Sigma_{ab}
           +     {\mbox{$1 \over 4$}}e^{\lambda}
		h^{cm}s^{n}\widetilde{\Sigma}_{mn}\Sigma_{ab}
           +   {\mbox{$1 \over 2$}}e^{\lambda}\widetilde{\Sigma}_{ab}\left(
	\mbox{$1 \over
		2$}h^{cm}s^{n}\Sigma_{mn}
           +     e^{-\lambda}w^{c} -\mbox{$1 \over
		2$}e^{-\lambda}\triangle^{c}\lambda\right)
		\nonumber \\
           &&\mbox{\ \ \ }-   {\mbox{$1 \over 4$}}e^{\lambda}s^{m}\triangle_m
	\lambda \, \Sigma_{ab}
           +     \mbox{$1 \over
		4$}e^{\lambda}s^{c}\Sigma_{nb}h^{nm}\widetilde{\Sigma}_{ma}
           -     {\mbox{$1 \over 4$}}
		e^{\lambda}s^{n}\widetilde{\Sigma}_{an}h^{cm}\Sigma_{bm}
	\end{eqnarray}

\subsection{The Weyl Scalars}\label{weyl}
We choose a complex-null tetrad $\{ e^{\lambda} \mbox{\boldmath
$l$\unboldmath},
\mbox{\boldmath $n$\unboldmath}, \mbox{\boldmath $m$\unboldmath},
\overline{\mbox{\boldmath $m$\unboldmath}} \} $ such that $ 2 m^{(\mu}
\overline{m}^{\nu)
}  =  h^{\mu \nu} $ and, \mbox{\boldmath
$l$\unboldmath} and \mbox{\boldmath
$n$\unboldmath} are given by (\ref{3.8}).
In this tetrad the five Newman-Penrose Weyl scalars are in a Ricci flat
spacetime
	\begin{eqnarray}
	\Psi_{0} =&& {\mbox{$1 \over 4$}} e^{2 \lambda} (2 {\cal L}_{l}\,
	\Sigma_{ab} + 2\nu
	\Sigma_{ab} -
           \Sigma_{am} h^{mn} \Sigma_{bn}) m^{a} m^{b}  \label{psi0}\\
	\Psi_{1} =&& {\mbox{$1 \over 4$}}e^{\lambda} ( -2 \triangle_{a}\nu +2
	\omega^{m}
	\Sigma_{am} +
            4 {\cal L}_{l}\, \omega_{a}
          + \triangle^{m} \lambda \Sigma_{ma}) m^{a} \\
	\Psi_{2} =&& -{\mbox{$1 \over 4$}} (2 e^{\lambda} {\cal L}_{n}\,
	\Sigma_{ab} + 2
	\triangle_{a}
             \triangle_{b} \lambda - \triangle_{a} \lambda \triangle_{b}
	\lambda
         - 4 \triangle_{b} \omega_{a} - e^{\lambda} \Sigma_{mb} h^{mn}
	\widetilde{\Sigma}_{an} +
             2\omega_{a} \triangle_{b} \lambda \nonumber \\
         && \mbox{\ \ \ \ \ }+ 2 \omega_{b} \triangle_{a} \lambda -
             4 \omega_{a} \omega_{b} +e^{\lambda}\triangle_{b} \lambda s^{m}
	\Sigma_{am}
         + 2 e^{\lambda} s^{m} \Sigma_{am} \omega_{b}) m^a \bar{m}^{b}  \\
	\Psi_{3} =&&  {\mbox{$1 \over 4$} } ( - 4 {\cal L}_{n}\, \omega_{a} - 2
	\triangle_{a}
	\widetilde{\nu} +
              e^{\lambda} s^{m}\Sigma_{mn}s^{n} \triangle_{a}\lambda
         + \triangle^{m} \lambda
              \widetilde{\Sigma}_{ma} + 2 e^{\lambda} s^{m}\Sigma_{mn}
	s^{n} \omega_{a}
         -  2 \omega^{m} \widetilde{\Sigma}_{ma})\bar{m}^{a} \\
	\Psi_{4} =&& -{\mbox{$1 \over 4$} }( -2 \widetilde{\Sigma}_{ab}
	\widetilde{\nu} - 2 {
	{\cal L}_{n}\,} \widetilde{\Sigma}_{ab} +
            \widetilde{\Sigma}_{am} h^{mn} \widetilde{\Sigma}_{bn}) \bar{m}^{a}
	\bar{m}^{b}
	\end{eqnarray}

\end{document}